\PassOptionsToClass{superscriptaddress}{revtex4-2}

\documentclass[fleqn,aps, prb, twocolumn,showpacs,floatfix,longbibliography]{revtex4-2}

\usepackage{amsmath,amssymb,amsfonts,float,graphics,epsfig,epstopdf,color,verbatim,tabularx,bm,multirow,appendix,hyperref}

\usepackage{amsthm}
\theoremstyle{definition}

\usepackage[normalem]{ulem} 
\usepackage{amsmath}
\usepackage{amssymb}
\usepackage{mathtools}
\usepackage{graphicx}
\usepackage{lipsum}
\usepackage{bm}
\usepackage{hyperref}
\usepackage{url}
\usepackage[utf8]{inputenc}
\usepackage{subfigure}
\usepackage{slashed,bbm}
\usepackage{graphics,psfrag,epsfig}
\usepackage{dsfont}
\usepackage{setspace}
\usepackage{wasysym}
\usepackage{slashed}
\usepackage{lipsum}
\usepackage{braket}
\usepackage{physics}
\usepackage{enumerate}
\usepackage{makecell}
\usepackage{xcolor}
\usepackage{ulem}
\usepackage{tikz}

\definecolor{dark-red}{rgb}{0.4,0.15,0.15}
\definecolor{dark-blue}{rgb}{0.15,0.15,0.4}
\definecolor{medium-blue}{rgb}{0,0,0.5}
\hypersetup{
colorlinks, linkcolor={dark-blue},
citecolor={dark-blue}, urlcolor={medium-blue}
}

\newcommand{\be}{\begin{equation}}
\newcommand{\ee}{\end{equation}}
\newcommand{\bea}{\begin{eqnarray}}
\newcommand{\eea}{\end{eqnarray}}

\renewcommand{\i}{\text{i}}

\begin{document}

\title{
    Markov length can diverge in systems whose universal physics is  spatially Markovian   
}
 
\author{Yu-Hsueh Chen} 
\affiliation{Department of Physics, University of California at San Diego, La Jolla, California 92093, USA} 
\affiliation{Kavli Institute for Theoretical Physics, University of California, Santa Barbara, California 93106, USA}
\author{Tarun Grover} 
\affiliation{Department of Physics, University of California at San Diego, La Jolla, California 92093, USA}

\begin{abstract}
Conditional mutual information (CMI) probes spatial non-Markovianity and plays a central role in a recently developed framework for defining mixed-state phases based on local reversibility. This framework sharply distinguishes exponentially decaying from algebraically decaying CMI, since the latter obstructs equivalence to a state with finite Markov length. Here we show that states that flow to the same
renormalization-group (RG) fixed point can nevertheless exhibit qualitatively different CMI. Starting from a critical Gibbs state of a finite-range local classical Hamiltonian,
whose CMI vanishes beyond the interaction range, we show that an RG-irrelevant perturbation that breaks
detailed balance generically generates algebraically decaying CMI.  %
{Crucially, these power-law tails are cutoff-suppressed: although they lead to an infinite Markov length at every fixed lattice cutoff, they vanish as a positive power of the cutoff in the continuum limit.}
We construct solvable models that exhibit this phenomenon and predict that the Ising-symmetric critical point of Toom's cellular automata generically exhibits such behavior. We also find analogous cutoff-suppressed CMI in the high-temperature paramagnetic phase of the long-range Ising paramagnet, in contrast with the genuine power-law CMI at the finite-temperature critical point in this same  model that survives the continuum limit. We also derive a general result that in one spatial dimension, CMI decays faster than the inverse square of the buffer length if and only if the long-distance physics is Markovian.

\end{abstract}
\maketitle

\textbf{Introduction.} 
The renormalization group (RG) allows one to define phases and phase transitions as equivalence classes of states that lie in the basin of attraction of distinct RG fixed points. States that flow to the same fixed point are expected to then share the same long-distance and long-time properties (``universal properties'') as the corresponding RG fixed point. For example, if one perturbs an RG fixed point with an irrelevant perturbation, then the leading scaling behavior of correlation functions is expected to remain unchanged. In this paper we will show that this expectation requires refinement for \textit{conditional correlations}, such as conditional mutual information (CMI). In particular, if one perturbs a dynamical RG fixed point whose steady state is a Gibbs state of a local Hamiltonian (`local Gibbs' for short) with power-law correlations, then we find that an \textit{irrelevant} detailed-balance-breaking perturbation generically causes the CMI, which is initially zero~\cite{clifford1971markov,leifer2008quantum,brown2012quantum}, to acquire a power-law decay. Crucially, the nature of such power laws is quite different from the power laws of ordinary correlation functions: they are suppressed by powers of the lattice cutoff and disappear in the scaling limit, where the ratio of the lattice cutoff to all macroscopic
length scales is sent to zero, thereby recovering the
spatially Markovian behavior of the fixed point.
This fingerprint allows one to distinguish such unusual power-law CMI from CMI that arises in systems such as 1+1-D directed percolation criticality, where the effective description of the system is genuinely non-Markovian and where the power-law CMI does not vanish as the lattice cutoff is sent to zero~\cite{chen2025local}.

\begin{figure}[t]
	\centering 
	\begin{tikzpicture}[scale=0.7, line cap=round, line join=round, >=stealth,
	every node/.style={font=\footnotesize, inner sep=1.5pt}]
	
	\node[font=\footnotesize\bfseries] at (-4.7,3.45) {(a)};
	\fill[gray!25]   (-3.4,-3.0) rectangle (3.4,3.0);
	\fill[blue!12]   (-2.3,-2.3) rectangle (2.3,2.3);
	\fill[orange!18] (-1.0,-1.0) rectangle (1.0,1.0);
	
	\draw[thick] (-2.3,-2.3) rectangle (2.3,2.3);
	\draw[thick] (-1.0,-1.0) rectangle (1.0,1.0);
	\draw[dashed] (-3.4,-3.0) rectangle (3.4,3.0);
	
	\node at (0,0) {$A$};
	\node at (0,1.65) {$B$};
	\node at (2.85,-2.65) {$C$};
	
	\draw[gray, thin] (-2.3,2.3) -- (-2.3,3.35);
	\draw[gray, thin] ( 2.3,2.3) -- ( 2.3,3.35);
	\draw[<->] (-2.3,3.25) -- (2.3,3.25) node[midway, above] {$l+2l_B$};
	
	\draw[<->] (-1.0,-1.45) -- (1.0,-1.45) node[midway, below] {$l$};
	\draw[<->] (1.0,0) -- (2.3,0) node[midway, above] {$l_B$};
	
	\begin{scope}[yshift=6mm]
	\node[font=\footnotesize\bfseries] at (-4.7,-4.15) {(b)};
	\draw[gray,thin,->] (-3.8,-4.75) -- (-3.2,-4.75);
	\draw[gray,thin,->] (3.2,-4.75) -- (3.8,-4.75);
	\fill[orange!18] (-3.2,-4.83) rectangle (-2.0,-4.67);
	\fill[blue!12] (-2.0,-4.83) rectangle (2.0,-4.67);
	\fill[orange!18] (2.0,-4.83) rectangle (3.2,-4.67);
	\draw[thick] (-3.2,-4.83) rectangle (3.2,-4.67);
	\draw[thick] (-2.0,-4.83) -- (-2.0,-4.67);
	\draw[thick] (2.0,-4.83) -- (2.0,-4.67);
	\node at (-2.6,-4.35) {$A$};
	\node at (0,-4.35) {$B$};
	\node at (2.6,-4.35) {$C$};
	\draw[<->] (-3.2,-5.25) -- (-2.0,-5.25) node[midway,below] {$l$};
	\draw[<->] (-2.0,-5.25) -- (2.0,-5.25) node[midway,below] {$l_B$};
	\draw[<->] (2.0,-5.25) -- (3.2,-5.25) node[midway,below] {$l$};
	\end{scope}
	
	\end{tikzpicture}
	
	\caption{Two geometries for the CMI $I(A:C|B)$.
		(a) The annular geometry, drawn as a two-dimensional cross-section: $ABC$ is the full infinite system, $A$ is a
		hypercube of side $l$, $A\cup B$ is a hypercube of side $l+2l_B$, so that
		$B$ is a shell of thickness $l_B$, and $C=\overline{A\cup B}$ extends to
		infinity (dashed outline).
		(b) In one dimension, $A$ and $C$ are intervals of length $l$ separated by an interval $B$ of length $l_B$.
	}
	\label{fig:geometries}
\end{figure}
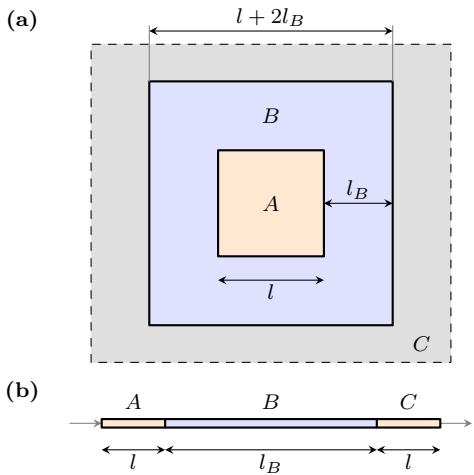

To set up the problem, consider a $d$-dimensional system partitioned into regions $A$, $B$, and $C$, with $B$ shielding $A$ from $C$. We will primarily use the
annular geometry of Ref.~\cite{sang2025stability}: $A$ is
a hypercube of linear size $l$, $A\cup B$ has
linear size $l+2l_B$, and
$C=\overline{A\cup B}$ [see Fig.~\ref{fig:geometries}(a)]. The CMI $I(A:C|B) = S(AB) + S(BC) - S(ABC) - S(B)$, where $S(X)$ is the von Neumann entropy of the state in region $X$, measures correlations between $A$
and $C$ that are not mediated by $B$.  
The Markov length $\xi_M$ is defined through $\textrm{CMI}\sim e^{-l_B/\xi_M}$ in the large-$l_B$ limit. Therefore, any subexponential decay in
$l_B$ corresponds to $\xi_M=\infty$.

The distinction between exponentially decaying and algebraic CMI plays a crucial role in defining mixed-state phases of matter~\cite{coser2019classification,rakovszky2024defining,sang2024mixed,sang2025stability,sang2025mixed}. As discussed in Refs.\cite{sang2025stability,sang2025mixed}, if a density matrix $\rho_1$ can be converted to another density matrix $\rho_2$ via a short-time local Lindblad evolution such that CMI continues to decay exponentially throughout the evolution, then it is natural to regard them as belonging to the same phase of matter since the exponential decay of CMI implies that one can find another short-time local Lindblad evolution that converts $\rho_2$ to $\rho_1$. Further, if $\rho_1$ has power-law CMI while $\rho_2$ has exponentially decaying CMI, then there cannot be a locally reversible short-depth channel connecting the two states~\cite{jinmin2026universal}.

However, as we alluded to above, not all power-law CMI has the same status. We distinguish
\emph{continuum-surviving} power-law CMI, which remains
finite in the continuum scaling limit, from
\emph{cutoff-suppressed} power-law CMI, whose amplitude
vanishes as a positive power of the lattice cutoff $a$. To make this distinction precise, let's assume that we are close to a scale-invariant fixed point so that there exists a crossover length scale $\xi_{\rm cross} \sim a |p|^{-1/\omega}$ defined by the tuning parameter $p$ with RG eigenvalue $\omega$. 
For a relevant perturbation, $\omega > 0$,  this length scale will approach infinity, i.e., $\xi_{\rm cross } \gg l, l_B$,
	 whereas for an irrelevant perturbation, $\omega < 0$, which is the case of interest here, it remains of the order of the lattice spacing \footnote{We emphasize that, in general, $\xi_M \neq \xi_{\rm cross}$ ($\xi_M$ can be understood as a ``derived'' quantity from $\xi_{\rm cross}$ and CMI), and our main focus is precisely the case in which $\xi_M$ diverges while $\xi_{\rm cross} \sim a$.}.
This implies that there are only three relevant length scales: $l$, $l_B$, and $a$. 
Since CMI is dimensionless, one must have  
$\textrm{CMI}(l_B,l,a)=\textrm{CMI}(l_B/l,l_B/a)$.   
The continuum-surviving power-law CMI remains non-zero as $a \to 0$ at any fixed $l_B/l$. In contrast, if CMI decays algebraically with $l_B/a$ as well, the resulting power-law tail is cutoff-suppressed and vanishes in this limit. 
Nevertheless, at any fixed $a$ it remains algebraic in $l_B$ and therefore corresponds to an infinite Markov length. Thus, despite disappearing in the continuum scaling limit,
cutoff-suppressed CMI remains an obstruction to local 
reversibility at any fixed microscopic cutoff \footnote{More concretely, for a lattice-regularized system, the continuum limit can be realized by taking
the dimensionless lengths $n=l/a$ and
$n_B=l_B/a$ to infinity at fixed ratio
$y=n_B/n$. Thus, a contribution that decays only
algebraically with $n_B$ remains a power law at every
finite lattice spacing, despite vanishing in the continuum
limit.}.
This feature may be understood as $\xi_{\rm cross} \sim a $ being a ``dangerously irrelevant'' parameter for CMI. 
A similar dangerously irrelevant nature for ordinary (non-conditional) correlations in nonequilibrium systems has also been pointed out recently in Ref.\cite{metzger2026generic}.
We note that the above argument relies on  CMI being UV-finite, i.e., $\lim_{a\to 0} \text{CMI} < \infty$, which we  assume throughout the paper \footnote{This is known to be true for many ground states of local quantum field theories \cite{fursaev2006entanglement,ryu2006aspects,solodukhin2008entanglement,metlitski2009,Grover11_2,liu2013refinement,casini2015mutual,van2025finite,Holzhey94,calabrese2004entanglement,Levin06,vardhan2024petz,balasubramanian2019information}, and has also been verified for several nonequilibrium systems
\cite{chen2025local,chen2026constraints,chen2026renormalization}. It is also consistent with the numerical observation in Ref.~\cite{yang2025topological} that the CMI at the critical point of the dephased toric code is scale invariant.
We point out that a nonlocal spacetime action, such as the long-range Ising model considered here, does \emph{not} obstruct UV-finite CMI: a mixed state $\rho$ has UV-finite CMI if it admits a locally reversible RG with a fixed
on-site Hilbert-space dimension~\cite{chen2026renormalization}.}. As an aside, for a relevant perturbation, one can also consider the regime where $l \gg l_B, \xi_{\rm cross}$, and as discussed recently~\cite{chen2026constraints}, in this case an information-theoretic inequality for CMI provides a non-perturbative constraint on the RG flow.

As we will discuss in detail below, a simple setting where
the aforementioned cutoff-suppressed CMI arises is Langevin dynamics whose unperturbed steady state is a critical Gibbs distribution of a finite-range local classical Hamiltonian, with power-law correlations
of local observables and power-law mutual information (MI)
between distant regions. The Markov property of such Gibbs distributions implies that $I(A:C|B)=0$ whenever $l_B$ is greater than the interaction range~\cite{clifford1971markov,leifer2008quantum,brown2012quantum}. We show that an irrelevant detailed-balance-breaking
perturbation of the dynamics generically produces non-universal
CMI with algebraic spatial decay. We further show that
the same distinction between genuine and cutoff-suppressed
power laws appears in the long-range Ising model. Here the critical point has genuine power-law CMI that survives the continuum limit, whereas the high-temperature paramagnetic phase, which flows to the infinite-temperature (trivial) state under RG \cite{fisher1972critical}, supports a cutoff-suppressed power-law CMI.

\textbf{One-dimensional CMI tails.} 
{We now present a simple yet powerful result for the scaling of CMI in one-dimensional systems.} 
Consider a translationally invariant one-dimensional mixed state $\rho$, which can be either classical or quantum, of infinite size close to the RG fixed point. 
Take $A$ and $C$ to be intervals of length $l$ separated by an interval $B$ of length $l_B$ [Fig.~\ref{fig:geometries}(b)].
{We now show that the CMI in this geometry decays faster than $l_B^{-2}$ if and only if the CMI vanishes as {the lattice cutoff} $a \to 0$.
This property, namely, $\lim_{a \to 0}\rm CMI = 0$, is what we mean by a fixed point being Markovian.
}

The essential point that leads to the claim is that, due to translational invariance, the CMI in Fig.\ref{fig:geometries}(b) is uniquely determined by the von Neumann entropy of an open interval $S(n)$ as $I(A:C|B) = 2S(n_B+n) - S(n_B) - S(n_B+2n)$, where we've introduced dimensionless variables $n_B = l_B/a$, $n = l/a$. 
A simple expansion when $n_B \gg n$  yields
$\text{CMI}(n,n_B) \simeq -n^2 d^2 S(n_B)/d n_B^2 = (n/n_B)^2 \times [-n_B^2 d^2 S(n_B)/d n_B^2]$ 
\footnote{We assume that the relative error is $o(1)$ in the joint limit $n_B\to\infty$ and $n/n_B\to0$.}
.
Reexpressing everything using  dimensionful variables, one finds that
\begin{equation}
\label{Eq:1d_cmi}
\textrm{CMI}(l,l_B, a)\simeq \Big(\frac{l}{l_B}\Big)^2 g\Big(\frac{l_B}{a}\Big).
\end{equation}
Therefore, the CMI (at fixed $l$ and $a$) decays faster than $l_B^{-2}$ if and only if $g(x)\to0$ as $x\to\infty$, which is  equivalent to the CMI vanishing as $a\to0$ at fixed $l$ and $l_B$.
This proves our aforementioned claim.
As an example, if $g(x) \sim x^{-\delta}$ with $\delta > 0$, one will obtain cutoff-suppressed CMI that decays as $l_B^{-(2+\delta)}$. 
We will later show that the high-temperature 1d long-range classical Ising model has such a scaling.

The above result constrains CMI of fixed points whose long-distance physics is spatially Markovian. Non-Markovian fixed points can  of course have CMI that does not satisfy this constraint. In particular, if CMI is UV finite and not cutoff-suppressed, i.e., it neither diverges nor vanishes as $a \to 0$, then  CMI $\sim (l/l_B)^2$. This conclusion has also been derived in Ref.~\cite{artiaco2025universal} assuming CMI is only a function of $l/l_B$. A few examples of such behavior are $(1+1)$-D conformal field 
theories, $(1+1)$-D directed percolation~\cite{chen2025local}, and a one-dimensional classical symmetry-breaking state under local decoherence studied in Refs.~\cite{lloyd2025diverging,chen2026constraints}.

Finally, the same argument implies that any power-law exponent $\eta<2$ would imply that $g(x)$ diverges when $x \to \infty$, and thus the CMI is not UV-finite.   
One example of this kind is the $s$-colored Motzkin chain~\cite{bravyi2012criticality,movassagh2016supercritical}, whose entanglement entropy for $s>1$ scales as $S(L)\sim L^{1/2}$~\cite{movassagh2016supercritical,movassagh2017entanglement}, implying $\eta=3/2$.

\textbf{{Cutoff-suppressed power-law} CMI from perturbing a critical Gibbs state.}  
We now present our main result: a generic detailed-balance-breaking perturbation of Langevin dynamics whose unperturbed steady state is a scale-invariant critical local Gibbs state induces a {cutoff-suppressed power-law CMI}.
 
Specifically, let $\phi(x,t)$ denote the field (or a collection of fields) and take
\begin{equation}
\label{Eq:langevin}
\partial_t \phi(x,t)=-\frac{\delta H [\phi]}{\delta\phi}+\epsilon\, v[\phi]+\zeta(x,t),
\end{equation}
where $H[\phi]$ is the local Hamiltonian, $\zeta(x,t)$ is Gaussian white noise with $\langle\zeta(x,t)\zeta(x',t')\rangle=2\delta^{d}(x-x')\delta(t-t')$, and $\epsilon v[\phi]$ is the perturbation that breaks detailed balance.
At $\epsilon=0$ the dynamics relaxes to the Gibbs distribution $P_0=e^{-H}/Z$. 
To compute the CMI to leading order in $\epsilon$, we combine the quadratic expansion of the CMI around a classical Markov distribution~\cite{kullback1962tests,kubkowski2021power} with the first-order McLennan--Zubarev expression for the steady state~\cite{mclennan1959statistical,zubarev1974nonequilibrium,maes2010rigorous}.
To state the result, let $X$ be any region with complement $\bar{X}$ and define $\langle O\rangle_{0|X}[\phi_X]
\equiv {\int\mathcal D\phi_{\bar X}\,e^{-H[\phi]}O[\phi]}/
{\int\mathcal D\phi_{\bar X}\,e^{-H[\phi]}}$, $\langle O\rangle_0\equiv\langle O\rangle_{0|\varnothing}$, and  $O_\times\equiv O-O_{AB}-O_{BC}+O_B$, with $O_X\equiv\langle O\rangle_{0|X}$. {Thus $O_X$ is the equilibrium expectation of $O$ conditioned on the field configuration in $X$, while $O_\times$ isolates the part of $O$ that cannot be captured by $AB$ and $BC$ separately.}
With this notation, we find that at the leading non-zero order in $\epsilon$ (see App.~\ref{app:perturbative_cmi} for details),
\begin{equation} 
\label{Eq:cmi_susceptibility} 
I_\epsilon(A:C|B)=\frac{\epsilon^2}{2}\int_0^\infty\!\!dt\int_0^\infty\!\!dt'\;
\big\langle (e^{t\mathcal{L}^{(0)}}\Sigma)_\times(e^{t'\mathcal{L}^{(0)}}\Sigma)_\times\big\rangle_0,
\end{equation}
where $\mathcal{L}^{(0)}=\int d^dx\left[\frac{\delta^2}{\delta\phi(\mathbf{x})^2}
-\frac{\delta H}{\delta\phi(\mathbf{x})}\frac{\delta}{\delta\phi(\mathbf{x})}\right]$ is the backward generator of the equilibrium dynamics and $\Sigma[\phi]=\int d^dx\left[v(\mathbf{x})\frac{\delta H}{\delta\phi(\mathbf{x})}
-\frac{\delta v(\mathbf{x})}{\delta\phi(\mathbf{x})}\right]$ is the source generated by the perturbation.   Although our derivation above used additive white noise, Eq.~\eqref{Eq:cmi_susceptibility} holds more generally for local noise that preserves detailed balance with $P_0$, upon replacing $\mathcal{L}^{(0)}$ by the appropriate unperturbed generator.

Using Eq.\eqref{Eq:cmi_susceptibility}, we now argue that a generic detailed-balance-breaking perturbation from a critical local Gibbs state with algebraically decaying mutual information will induce algebraically decaying CMI.

Being an irrelevant perturbation implies that $\xi_{\rm cross} =  \epsilon^{1/|\omega|}$ (and hence  $\epsilon^2  = \xi_{\rm cross}^{2 |\omega|}$) with  $\omega <0$ the RG eigenvalue. 
On the other hand, the double integrals in Eq.\eqref{Eq:cmi_susceptibility} are completely determined by the {unperturbed} local Gibbs state, which we assume to be scale invariant, and thus the double integrals can only depend on the shape of the tripartition and three length scales $l_B$, $l$, and $a$.
In the scaling limit $l, l_B \gg a$, one then finds
\begin{equation} 
\label{Eq:cmi_critical}
\text{CMI}={\frac{1}{2}}\left(\frac{\xi_{\rm cross}}{l_B}\right)^{2|\omega|} \Phi(l_B/l),
\end{equation}
where $\Phi(l_B/l) = l_B^{2|\omega|}\int_0^\infty\!\!dt\int_0^\infty\!\!dt'\;
\big\langle (e^{t\mathcal{L}^{(0)}}\Sigma)_\times(e^{t'\mathcal{L}^{(0)}}\Sigma)_\times\big\rangle_0$ up to corrections that vanish  as $a \to 0$. 
One expects that for a generic detailed-balance-breaking perturbation, the operators $(e^{t\mathcal{L}^{(0)}}\Sigma)_\times$ have a nonzero overlap with scaling operators of the critical fixed point. Due to scale invariance, the corresponding correlators decay as powers of distance. Integrating these correlators over regions of size $l$ and $l_B$ then produces powers of their ratio, and thus $\Phi(y)$ is expected to decay as a power of $y$ in the $y \gg 1$ (i.e., $l_B\gg l$) limit. Therefore, the Markov length diverges.

{As a concrete application, consider the Ising-symmetric dynamical critical point of Toom's cellular automata~\cite{toom1974nonergodic,toom1980stable}.}
Following Ref.~\cite{he1990generic}, the Toom-like dynamics can be described by the Langevin equation of Eq.\eqref{Eq:langevin} with the Hamiltonian being the $\phi^4$ theory $H=\int d^2x\,[\tfrac{1}{2}(\nabla\phi)^2+\tfrac{1}{2}m^2\phi^2+\tfrac{u}{4!}\phi^4]$ and $v_{\rm toom}[\phi] = (\partial_x^3+\partial_y^3)\phi$.
Since $v_{\rm toom}[\phi]$ has three spatial derivatives, this perturbation is expected to be irrelevant at the unperturbed  dynamical fixed point, which corresponds to the standard model-A~\cite{grinstein1985statistical}. Using the scaling analysis of Ref.~\cite{shinkai1992bistability} together with the numerical results in Ref.~\cite{Godreche2017generic}, we estimate $\omega \approx -1.3 \pm 0.2$. 
We therefore predict that the Ising-symmetric critical point of Toom's cellular automaton exhibits cutoff-suppressed power-law CMI.

We note that while here we expand CMI only up to $O(\epsilon^2)$, essentially the same argument for a divergent Markov length goes through in higher-order expansions.

{Above we showed that a generic detailed-balance-breaking perturbation to the dynamical fixed point of a critical, {classical} Gibbs state leads to cutoff-suppressed power-law CMI. A useful sufficient condition for \textit{excluding} such CMI, at least at the leading order (Eq.\eqref{Eq:cmi_susceptibility}), is rapid mixing together with locality of the unperturbed dynamics.  By rapid mixing we mean that there is a system-size-independent relaxation rate $\gamma>0$ such that any zero-mean observable $O$ satisfies $\|e^{t\mathcal L^{(0)}}O\|_{2,0}\le e^{-\gamma t}\|O\|_{2,0}$, where $\|O\|_{2,0}^2\equiv\langle O^2\rangle_0$.  Locality implies that if $O_X$ is initially supported in a region $X$, then after time $t$ it can be approximated by an observable supported within distance $R$ of $X$, with an error that is exponentially small when $R$ is larger than the distance information can propagate in time $t$; schematically, $\|e^{t\mathcal L^{(0)}}O_X-O_{X_R}(t)\|_{2,0}\lesssim e^{vt-\mu R}$, where $O_{X_R}(t)$ is supported within distance $R$ of $X$, and $v,\mu>0$.  Applying this to the local terms in $\Sigma$ in Eq.~\eqref{Eq:cmi_susceptibility}, the $\times$ projection can only retain contributions that have propagated across the buffer, and therefore $\|(e^{t\mathcal L^{(0)}}\Sigma)_\times\|_{2,0}\lesssim\min\{e^{-\gamma t},e^{vt-\mu l_B}\}$.  Integrating over $t$ in $V=\int_0^\infty dt\,e^{t\mathcal L^{(0)}}\Sigma$ then gives $\|V_\times\|_{2,0}\lesssim e^{-\kappa l_B}$ with $\kappa=\mu\gamma/(v+\gamma)>0$.  Hence Eq.~\eqref{Eq:cmi_susceptibility} gives $I(A:C|B)\lesssim \epsilon^2 e^{-2\kappa l_B}$ at leading order. This condition also suggests mechanisms other than the critical slowing down for generating cutoff-suppressed power-law CMI. For example, conserved dynamics need not be rapidly mixing: even when the steady state itself has short-ranged correlations, long-wavelength fluctuations can relax diffusively, so the time integral in Eq.~\eqref{Eq:cmi_susceptibility} can generate cutoff-suppressed power-law mutual information (MI)~\cite{grinstein1991generic,damman2024algebraic,metzger2026generic} and CMI.  We give an explicit example below.}

\textbf{General CMI scaling for Gaussian systems.}
Above we derived a perturbative expression for cutoff-suppressed power-law CMI in critical Gibbs states perturbed by a detailed-balance-breaking term. Since this is a general phenomenon, it is instructive to demonstrate it non-perturbatively in solvable models. To that end, we derive the scaling of the CMI for a broad class of Gaussian theories in which a local critical kernel is supplemented by an irrelevant nonlocal contribution.
Let us first recall that a translationally invariant Gaussian distribution is fully specified by its quadratic kernel $K(\mathbf q)$ as 
$P_{}[\phi]\propto\exp[-\frac{1}{2}\int_{\mathbf q\neq0}
\phi(\mathbf q)K(\mathbf q)\phi(-\mathbf q)]$, 
where $\int_{\mathbf q}\equiv\int d^d q/(2\pi)^d$.
Being a nonlocal Gibbs distribution here simply means that $K(\mathbf q)$ is nonanalytic as $\mathbf q \to 0$, as local terms are finite-derivative contact 
operators and thus are all analytic after Fourier transformation.

Consider the scenario where $K$ can be decomposed into a local scale-invariant part $K^{\rm loc}$ with $K^{\rm loc}(\lambda\mathbf q)=\lambda^{h_{\rm loc}}K^{\rm loc}(\mathbf q)$ and an irrelevant nonlocal part with $K^{\rm nloc}(\lambda\mathbf q)=\lambda^{h_{\rm nloc}}K^{\rm nloc}(\mathbf q)$.
Specifically,
\begin{equation}
\label{Eq:precision_two_term}
K(\mathbf q;\xi_{\rm cross})=K^{\rm loc}(\mathbf q)
+\xi_{\rm cross}^{|\omega|}K^{\rm nloc}(\mathbf q),
\end{equation}
where $\omega = h_{\rm loc} - h_{\rm nloc}$ is the scaling dimension of the nonlocal fixed point, which we assume to be \emph{irrelevant}, i.e., $\omega = -|\omega|<0$.
For a local critical Gaussian fixed point, $h_{\rm loc}=2$, which comes from the $\nabla^2$ term in the action. 
We note that $\xi_{\rm cross}$ in Eq.\eqref{Eq:precision_two_term} ensures that $K^{\rm loc}$ and $\xi_{\rm cross}^{|\omega|} K^{\rm nloc}$ carry the same dimension. 

In the End Matter, we present a simple scaling argument for the CMI in the geometry of Fig.~\ref{fig:geometries}(a), independent of further details of $K^{\rm loc}$ and $K^{\rm nloc}$. 
For $l_B \gg l$, the argument gives the following CMI scaling forms in three regimes distinguished by their relation to $\xi_{\rm cross}$: 
\begin{enumerate}[(i)]
    \item When $\xi_{\rm cross}\ll l\ll l_B$, which is the regime where Eq.\eqref{Eq:cmi_susceptibility} applies, one finds
    \begin{equation}
\label{Eq:large_y_regime_a}
\text{CMI} \sim \Big(\frac{\xi_{\rm cross}}{l_B } \Big)^{2|\omega|}  \Big(\frac{l}{l_B } \Big)^{d+h_{\rm loc}}.
\end{equation}
This supports the perturbative result in Eq.\eqref{Eq:cmi_critical}, where the $\xi_{\rm cross}/l_B$ dependence is determined by the irrelevant eigenvalue while the $l/l_B$ dependence is determined by the local theory.
    \item When $l\ll\xi_{\rm cross}\ll l_B$, one finds that the exponent for $l_B/l$ now depends on the \emph{nonlocal} scale-invariant theory as
\begin{equation}
\label{Eq:large_y_regime_b}
\text{CMI} \sim \Big(\frac{\xi_{\rm cross}}{l_B } \Big)^{|\omega|}\,\Big(\frac{l}{l_B } \Big)^{d+h_{\rm nloc}} .
\end{equation}
\item Finally, when $l\ll l_B\ll\xi_{\rm cross}$, one finds
\begin{equation}
\label{Eq:large_y_regime_c}
\text{CMI} \sim \Big(\frac{l}{l_B } \Big)^{d+h_{\rm nloc}},
\end{equation}
which is consistent with the intuition that if $\xi_{\rm cross}$ is greater than all the length scales at which we probe the system, then one will observe the scale-invariant behavior associated with the non-local kernel (``UV fixed point'').
\end{enumerate}

\textbf{An exactly solvable two-component Toom-like model.}
As a concrete example, we now discuss a Gaussian model perturbed with the Toom-like detailed-balance-breaking term $\mathcal{D}_3 \phi = (\partial_x^3+\partial_y^3)\phi$.

To keep the state Gaussian while {rendering} $K(\mathbf q)$ nonanalytic, we introduce a second field and give the two fields opposite spatial anisotropies, which is a bit similar to the role of anisotropy played in Ref.\cite{grinstein1990conservation}.
Specifically, we collect two real fields in $d = 2$ as $\Phi=(\phi,\psi)$, and then consider the Langevin equation 
\begin{equation}
\label{Eq:two_component}
\begin{aligned}
& \partial_t\Phi  =-(S+A)\Phi+\zeta,\\
& S  =-\begin{pmatrix}\partial_x^2+c\partial_y^2&0\\0&c\partial_x^2+\partial_y^2\end{pmatrix},\
A =-\epsilon(\partial^3_x+\partial^3_y)\begin{pmatrix}0&1\\1&0\end{pmatrix},
\end{aligned}
\end{equation} 
with $c>0$, $c\neq1$, and $\zeta(x,t)$ Gaussian white noise with $\langle\zeta(x,t)\zeta(x',t')\rangle=2  \delta^{2}(x-x')\delta(t-t')$.

\begin{figure}[t]
\centering
\includegraphics[width=0.9\columnwidth]{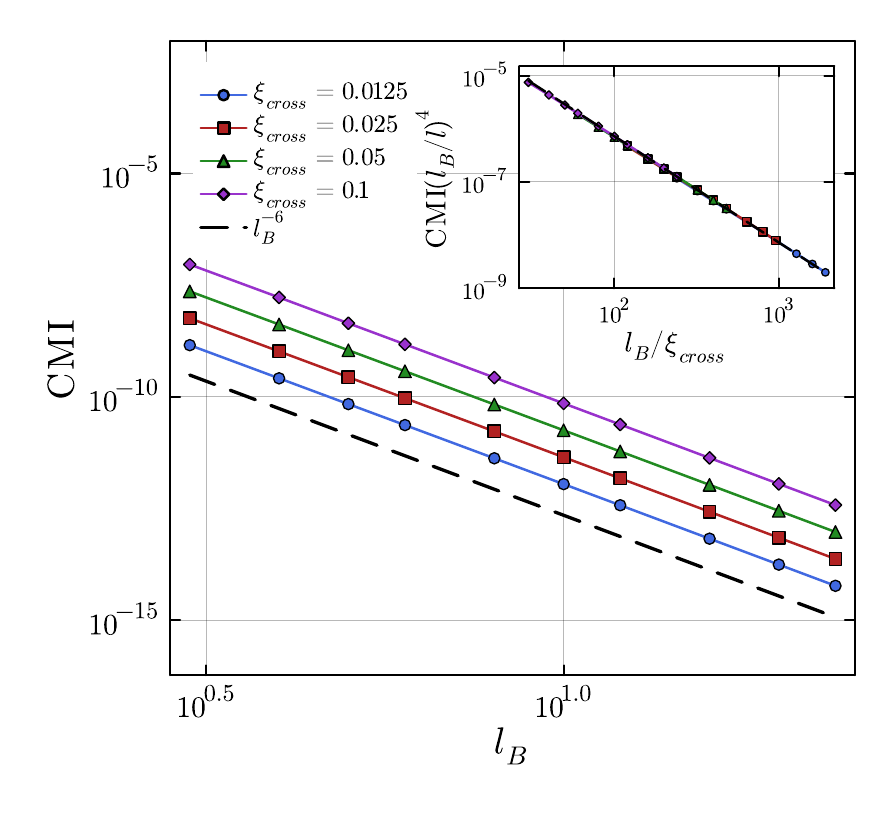}
\caption{
    CMI of the model in
Eq.~\eqref{Eq:two_component} on a periodic $128\times128$ lattice with
$c=2$ and a single-site region $A$ ($l=1$). All lengths are in lattice units. Main: CMI versus $l_B$ for several
$\xi_{\rm cross}=\epsilon$.
Inset: CMI$\,(l_B/l)^4$ versus $l_B/\xi_{\rm cross}$.   
}
\label{fig:two_component_numerics}
\end{figure}

Two {consequences} follow. First, $S \Phi = \delta H/\delta \Phi$ with $H$ the critical Gaussian
Hamiltonian $H=\frac12\int d^2x\,[(\partial_x\phi)^2+c(\partial_y\phi)^2+c(\partial_x\psi)^2+(\partial_y\psi)^2]$. Therefore, the steady state when $\epsilon = 0$ is a local Gibbs state that has infinite correlation length (but zero Markov length).
Second, the second term $A \Phi$ is antisymmetric, and thus it breaks detailed balance.  
At the Gaussian fixed point $z=2$, so $\omega = z-3 = -1$, and thus $\xi_{\rm cross}=\epsilon$. 
Since $h_{\rm loc} = 2$, without knowing any details about $K_{\rm nloc}$, one can already predict from Eq.\eqref{Eq:large_y_regime_a} that generically, $\text{CMI} \sim (\xi_{\rm cross}/{l_B})^{2 |\omega|} ({l}/{l_B})^{d+ h_{\text{loc}}} =(\xi_{\rm cross}/{l_B})^{2 }({l}/{l_B})^{4} $.

We now numerically verify the predicted dependence on $l_B$ and $\epsilon$ by evaluating the CMI with the geometry in Fig.\ref{fig:geometries}(a) from {the exact steady-state}. 
We set $c=2$, take $A$ to be a single site ($l=1$), and consider how the CMI scales as a function of $l_B$, measuring lengths in lattice units. In practice, one has to keep the whole system finite, so we consider a periodic $128\times128$ lattice.  
Fig.~\ref{fig:two_component_numerics} shows the predicted $l_B^{-6}$ decay 
and the collapse of $I(A:C|B)\,(l_B/l)^4$ versus $l_B/\xi_{\rm cross}$ onto
a curve of slope $-2$. 
Further numerical details are given in App.~\ref{app:two_component_exact}.
 
As hinted earlier, even when the unperturbed steady state has short-ranged correlations, slow relaxation dynamics can generate cutoff-suppressed power-law CMI. Keeping $S$ and $A$ in Eq.~\eqref{Eq:two_component} unchanged, consider
instead diagonal conserved noise with momentum-space covariance
$D(\mathbf k)=\mathrm{diag}\!\left[T_\phi(k_x^2+c k_y^2),
T_\psi(c k_x^2+k_y^2)\right]$, where $T_\phi, T_\psi$ are arbitrary noise strengths with $T_\phi \neq T_\psi$. At $\epsilon=0$ the stationary distribution is simply	$P_0[\phi,\psi]\propto
\exp\left[-\frac12\int d^2x
\left(\frac{\phi^2}{T_\phi}+\frac{\psi^2}{T_\psi}\right)\right]$ and thus fields at distinct spatial points are independent, and the MI and CMI {vanish identically}. Nevertheless, the same
Toom-like perturbation generates the off-diagonal elements of the quadratic kernel, 
$\delta K_{\phi\psi}(\mathbf k)
=
\frac{i\epsilon}{1+c}
\left(\frac{1}{T_\phi}-\frac{1}{T_\psi}\right)
\frac{k_x^3+k_y^3}{k_x^2+k_y^2}
+O(\epsilon^2)$,
and one may verify it produces cutoff-suppressed power-law CMI (as well as MI).

\textbf{Long-range Ising model.}
Having discussed a few examples of nonuniversal power-law tails, we close with a classic example where one can tune a parameter to access the crossover between universal and non-universal power-law tails: the 
long-range Ising model in $d$ spatial dimensions, with $\rho \propto \exp(-H)$ and the Hamiltonian   
\begin{equation}
\label{Eq:long_range_ising}
H=-\frac{\beta}{2}\sum_{i\neq j}\frac{s_i s_j}{|\mathbf{r}_i-\mathbf{r}_j|^{d+\sigma}},
\end{equation}
where $s_i = \pm 1$ denotes the Ising spin. 
We briefly discuss how this example illustrates the generic 1d result and the CMI scaling for Gaussian systems, with details collected in App.\ref{app:lr_ising}.

Let us first consider the irrelevant CMI tails.
For any $d$ and $\sigma>0$, for sufficiently small $\beta$, the long-range interaction is irrelevant and the system flows to a trivial product state \footnote{When $\sigma <0$, the aligned-state energy per spin diverges in the thermodynamic limit, implying that the Gibbs state at any $\beta>0$ is essentially $[(|0\rangle \langle 0|)^{N} + (|1\rangle \langle 1|)^{N}]/2 $, which has zero CMI.  
}. 
Therefore, our 1d result discussed in Eq.\eqref{Eq:1d_cmi} predicts that the CMI tail $l_B^{-\eta}$ with the geometry in Fig.\ref{fig:geometries}(b) has exponent  $\eta >2$.
Indeed, a standard high-temperature expansion in App.\ref{app:lr_ising} shows that $\eta = 2(1+\sigma)>2$ for any $\sigma>0$.

To make analytical progress close to the critical point, we consider the Gaussian regime $0<\sigma<d/2$ so that the physics is captured by a quadratic kernel of the form $K(\mathbf q)=\mu+|\mathbf{q}|^\sigma$, where $\mu = 0$ corresponds to the critical inverse temperature $\beta = \beta_c$ \cite{fisher1972critical}.  
When $d=1$, we find that the CMI in the geometry of Fig.~\ref{fig:geometries}(b) can be computed analytically at $\mu = 0$, and in the $l_B\gg l$ limit, we obtain $\text{CMI} = \sigma^2(l/l_B)^2/8+O((l/l_B)^3)$ (see App.~\ref{app:lr_ising}).
Therefore, the CMI has a universal tail with $\eta = 2$ that survives the continuum limit, reflecting the nonlocality of the critical Gaussian theory. 
{More generally, in any spatial dimension, Eqs.\eqref{Eq:large_y_regime_a}-\eqref{Eq:large_y_regime_c}, with $h_{\rm loc} = 0$, $h_{\rm nloc} = \sigma$ (and thus $|\omega| = |h_{\rm loc} - h_{\rm nloc}| =\sigma$), describe the crossover from genuine power-law CMI at the critical point ($\xi_{\rm cross }= \infty$) to cutoff-suppressed power-law CMI in the high-temperature phase at distances $l_B \gg \xi_{\rm cross }$.} 
We numerically verify Eqs.\eqref{Eq:large_y_regime_b}-\eqref{Eq:large_y_regime_c} in $d = 2$ (see App.~\ref{app:lr_ising}).

\textbf{Discussion.}
We have shown that the Markov length can diverge in systems whose universal physics is Markovian. 
{This exposes a tension between the conventional RG notion of universality, in which states flowing to the same long-distance fixed point are regarded as equivalent~\cite{cardy1996scaling}, and channel-based notions of mixed-state phase equivalence~\cite{sang2025stability,sang2025mixed,sang2024mixed}.
On that note, channels of polylogarithmic range cannot connect a state with power-law CMI to one with exponentially decaying CMI through a locally reversible path~\cite{jinmin2026universal}, 
whereas a concrete example shows that channels of sublinear range can~\cite{chen2026renormalization}.
It is then natural to wonder if one should relax the channel range in mixed-state phase equivalence from polylogarithmic~\cite{sang2025mixed} to sublinear~\cite{sang2024mixed} in the system size. 
 Alternatively, one may consider enlarging mixed-state phase equivalence to include short-depth channels with algebraic spatial tails, provided the strength of these long-range terms is itself cutoff-suppressed.}
 
One concrete prediction of our result is a cutoff-suppressed power-law CMI tail at the critical point on Toom's symmetric line, despite its expected Markovian universal physics~\cite{grinstein1985statistical,he1987numerical,Godreche2017generic}.
Notably, the exponent of such a power-law tail is universal and encodes the irrelevant RG eigenvalue. Apart from Toom's model, several cellular automata have been constructed that can stabilize classical memory without symmetry~\cite{pajouheshgar2026exploring,lake2025squeezing}. 
It would be interesting to study the CMI in these models, as their exponents differ from those of Model~A and may lead to the genuinely power-law CMI that survives the continuum limit.

Finally, we would like to highlight an upcoming work~\cite{negari2026}, which constructs one-dimensional examples where an arbitrarily weak local quantum channel generates power-law CMI, $I(A:C|B)\sim l_B^{-\eta}$ with $\eta>2$, and hence an infinite Markov length at any nonzero channel strength.  This provides another realization of
cutoff-suppressed power-law CMI in a rather different setting. 

\textbf{Acknowledgements.} 
We thank Tim Hsieh and Ethan Lake for helpful discussions.
We acknowledge use of ChatGPT 6 for helpful discussions and writing code.
T.G. is supported by the National Science Foundation under Grant No. DMR-2521369. 
This research was supported in part by the following grants to the Kavli Institute for Theoretical Physics (KITP): NSF PHY-2309135, Heising-Simons Foundation, and the Simons Foundation International SFI-MPS-T-Institutes-00021142, LB.

% Register references cited only inside a bibliography footnote on the first pass.
\nocite{fursaev2006entanglement,ryu2006aspects,solodukhin2008entanglement,metlitski2009,Grover11_2,liu2013refinement,casini2015mutual,van2025finite,Holzhey94,calabrese2004entanglement,Levin06,vardhan2024petz,balasubramanian2019information,yang2025topological}
\bibliography{bibs}

\clearpage
\twocolumngrid
\appendix
\section*{End Matter}

The goal of this End Matter is to present a simple argument that
yields the scaling exponents of the CMI $I(A:C|B)$ given in Eqs.~\eqref{Eq:large_y_regime_a}--\eqref{Eq:large_y_regime_c} in the main text, for a Gaussian distribution $P[\phi] \propto \exp[-\frac{1}{2}\sum_{i,j}\phi_i K_{i,j}\phi_j]$.
While we restrict ourselves to a system
partitioned into $A\cup B\cup C$ with the annular geometry in Fig.~\ref{fig:geometries}(a), the same argument extends naturally to other geometries. 
We therefore think it is valuable to present the simple argument here.

The CMI for a Gaussian distribution has been implicitly developed in Refs.~\cite{gelfand1959calculation,bach2002kernel,gretton2005kernel,barnett2009granger} and takes the form
\begin{equation}
\label{Eq:gaussian_cmi_trace_log}
\begin{aligned}
I(A:C|B)
&=-\frac{1}{2}\Tr\log(1-{\cal T})=\frac{1}{2}\sum_{m=1}^{\infty}\frac{1}{m}\Tr({\cal T}^m),\\
{\cal T}& \equiv
K_A^{-1}K_{A,C}K_C^{-1}K_{C,A}.
\end{aligned}
\end{equation}
Here, $K_X \equiv K_{X,X}$ is the diagonal block of the quadratic kernel $K$, while $K_{A,C}$ denotes the off-diagonal block of $K$.
In the continuum limit, one has
\begin{equation}
\label{Eq:bridge_continuum}
\begin{aligned}
  {\cal T}(\mathbf a,\mathbf a')& =\int_A d^d x\int_C d^d c\,d^d c'  \\
 & K_A^{-1}(\mathbf a,\mathbf x) 
  K_{A,C}(\mathbf x,\mathbf c)
K_C^{-1}(\mathbf c,\mathbf c')K_{C,A}(\mathbf c',\mathbf a').
\end{aligned}
\end{equation}

Recall that we consider the scenario where $K$ can be decomposed into a local scale-invariant part $K^{\rm loc}$ with $K^{\rm loc}(\lambda\mathbf q)=\lambda^{h_{\rm loc}}K^{\rm loc}(\mathbf q)$ and an irrelevant nonlocal part with $K^{\rm nloc}(\lambda\mathbf q)=\lambda^{h_{\rm nloc}}K^{\rm nloc}(\mathbf q)$, i.e.,
\begin{equation}
\label{Eq:precision_two_term_appendix}
\begin{aligned}
K(\mathbf q;\xi_{\rm cross})
&=K^{\rm loc}(\mathbf q)+\xi_{\rm cross}^{|\omega|}K^{\rm nloc}(\mathbf q),\end{aligned}
\end{equation}
where $\omega =h_{\rm loc}-h_{\rm nloc} = -|\omega|<0
$ so that the nonlocal part is irrelevant. 
This implies that  the local part dominates at lengths much
larger than $\xi_{\rm cross}$, whereas the nonlocal part dominates at lengths
much smaller than $\xi_{\rm cross}$.
In the following, we assume that the dominant part has a
well-defined restricted inverse and that the continuum integrals below converge.
 
In the $l_B \gg l$ limit, the leading term in Eq.~\eqref{Eq:gaussian_cmi_trace_log} is
\begin{equation}
\label{Eq:gaussian_cmi_large_y_leading}
\begin{aligned}
I(A:C|B)
&=\frac12\Tr\!\left[K_A^{-1}K_{A,C}K_C^{-1}K_{C,A}\right]+O\!\left(\Tr\mathcal T^2\right).
\end{aligned}
\end{equation}
Since local interactions do not contribute
to $K_{A,C}$ or $K_{C,A}$, only 
$\xi_{\rm cross}^{|\omega|}K^{\rm nloc}$ contributes to both of them.
Therefore, 
\begin{equation}
\label{Eq:large_y_explicit}
\begin{aligned}
&I(A:C|B)=\frac{\xi_{\rm cross}^{2|\omega|}}{2}
\int_A d^d a\,d^d a'\int_C d^d c\,d^d c'\\
&
K_A^{-1}(\mathbf a,\mathbf a';\xi_{\rm cross})
K^{\rm nloc}(\mathbf a'-\mathbf c)
K_C^{-1}(\mathbf c,\mathbf c';\xi_{\rm cross})
K^{\rm nloc}(\mathbf c'-\mathbf a),
\end{aligned}
\end{equation}
up to an $O(\Tr\mathcal T^2)$ correction.

The CMI scaling can now be deduced by simple counting of the length factors in the integral of Eq.\eqref{Eq:large_y_explicit}.
Let us first consider the part that is independent of $\xi_{\rm cross}$, i.e., $\int_A d^d a d^d a'$, $\int_C d^d c\,d^d c'$, $K^{\rm nloc}(\mathbf a'-\mathbf c)$,  and $K^{\rm nloc}(\mathbf c'-\mathbf a)$. 
The two integrations over
$A$ supply $l^{2d}$, and measuring the two coordinates in $C$ in units of
$l_B$ supplies $l_B^{2d}$. 
On the other hand, to determine the length scale each nonlocal kernel supplies, we recall that $K(\mathbf r)\sim \int {d^d q}e^{i\mathbf q\cdot\mathbf r}K(\mathbf q)$, and thus 
\begin{equation}
K^{\rm loc/nloc}(\lambda\mathbf r)
=\lambda^{-d-h_{\rm loc/nloc}}K^{\rm loc/nloc}(\mathbf r).
\end{equation}
This implies each nonlocal kernel supplies
$l_B^{-(d+h_{\rm nloc})}$.
More concretely, 
at fixed $\mathbf c/l_B$, one has 
\begin{equation}
\begin{aligned}
K^{\rm nloc}(\mathbf a'-\mathbf c)
&=l_B^{-(d+h_{\rm nloc})}
K^{\rm nloc}\!\left(-\frac{\mathbf c}{l_B}+\frac{\mathbf a'}{l_B}\right)\\
&=l_B^{-(d+h_{\rm nloc})}
\left[K^{\rm nloc}\!\left(-\frac{\mathbf c}{l_B}\right)
+O\!\left(\frac{l}{l_B}\right)\right].
\end{aligned}
\end{equation}
The same argument applies to $K^{\rm nloc}(\mathbf c'-\mathbf a)$.
Therefore, to leading order in $l/l_B$, both kernels are independent of the coordinates within $A$.
As a consequence, apart from $K_A^{-1}$ and $K_C^{-1}$, the length dependence is
\begin{equation}
\label{Eq:large_y_powers}
\xi_{\rm cross}^{2|\omega|}\ 
l^{2d}\ l_B^{2d} \ 
{l_B^{-2(d+h_{\rm nloc})}}= \xi_{\rm cross}^{2|\omega|}\ 
{l^{2d}}\ 
{l_B^{-2 h_{\rm nloc}}}.
\end{equation}

On the other hand, the length dependences of $K_A^{-1}$ and $K_C^{-1}$ depend on whichever
part of the kernel dominates at the relevant length. 
To find their scaling, we note that the restricted  inverse is defined by
$\int_X d^d x\,K(\mathbf r-\mathbf x;\xi_{\rm cross})K_X^{-1}(\mathbf x,\mathbf r';\xi_{\rm cross})=\delta^d(\mathbf r-\mathbf r')$
for $\mathbf r,\mathbf r'\in X$. 
Together with the scaling of $K$, i.e., $K(\lambda\mathbf r;\lambda\xi_{\rm cross})
=\lambda^{-d-h_{\rm loc}}K(\mathbf r;\xi_{\rm cross})$, this gives
\begin{equation}
\label{Eq:restricted_inverse_scaling}
\begin{aligned}
&\left(K_{\lambda X}^{-1}\right)
(\lambda\mathbf r,\lambda\mathbf r';\lambda\xi_{\rm cross})=\lambda^{h_{\rm loc}-d}
\left(K_X^{-1}\right)(\mathbf r,\mathbf r';\xi_{\rm cross}).
\end{aligned}
\end{equation}
In the local regime the inverse has spatial degree $h_{\rm loc}-d$.
In the nonlocal regime,
$K_X^{-1}\simeq\xi_{\rm cross}^{-|\omega|}(K_X^{\rm nloc})^{-1}$,
so its spatial degree is $h_{\rm nloc}-d$.  
It follows that
\begin{equation}
\label{Eq:restricted_inverse_regimes}
\begin{aligned}
K_A^{-1}&\sim
\begin{cases}
l^{h_{\rm loc}-d},&\xi_{\rm cross}\ll l,\\
\xi_{\rm cross}^{-|\omega|}l^{h_{\rm nloc}-d},&l\ll\xi_{\rm cross},
\end{cases}\\[3pt]
K_C^{-1}&\sim
\begin{cases}
l_B^{h_{\rm loc}-d},&\xi_{\rm cross}\ll l_B,\\
\xi_{\rm cross}^{-|\omega|}l_B^{h_{\rm nloc}-d},&l_B\ll\xi_{\rm cross}.
\end{cases}
\end{aligned}
\end{equation}
Here $\sim$ suppresses dimensionless coordinate dependence and shape factors.
Combining Eq.~\eqref{Eq:large_y_powers} with the two inverse factors in
Eq.~\eqref{Eq:restricted_inverse_regimes}, one finds:

\begin{enumerate}[(i)]
\item When $\xi_{\rm cross}\ll l\ll l_B$, the local part controls both
restricted inverses. Hence (recall $\omega =h_{\rm loc}-h_{\rm nloc} = -|\omega|$)
\begin{equation}
\label{Eq:large_y_regime_a_appendix}
\begin{aligned}
I(A:C|B)
&\sim (\xi_{\rm cross}^{2|\omega|}\ 
{l^{2d}}\ 
{l_B^{-2 h_{\rm nloc}}}) (l^{h_{\rm loc}-d}\ l_B^{h_{\rm loc}-d})\\
&=\left(\frac{\xi_{\rm cross}}{l_B}\right)^{2|\omega|}
\left(\frac{l}{l_B}\right)^{d+h_{\rm loc}}.
\end{aligned}
\end{equation}
\item When $l\ll\xi_{\rm cross}\ll l_B$, the nonlocal part controls
$K_A^{-1}$, while the local part still controls $K_C^{-1}$.
We therefore find that 
\begin{equation}
\label{Eq:large_y_regime_b_appendix}
\begin{aligned}
I(A:C|B) & \sim  (\xi_{\rm cross}^{2|\omega|}\ 
{l^{2d}}\ 
{l_B^{-2 h_{\rm nloc}}})(\xi_{\rm cross}^{-|\omega|}l^{h_{\rm nloc}-d} l_B^{h_{\rm loc}-d}) \\
&=
\left(\frac{\xi_{\rm cross}}{l_B}\right)^{|\omega|}
\left(\frac{l}{l_B}\right)^{d+h_{\rm nloc}}.
\end{aligned}
\end{equation}
\item When $l\ll l_B\ll\xi_{\rm cross}$, the nonlocal part controls both
restricted inverses at the relevant distances, and thus
\begin{equation}
  \begin{aligned}
\label{Eq:large_y_regime_c_appendix}
I(A:C|B) & \sim  (\xi_{\rm cross}^{2|\omega|}\ 
{l^{2d}}\ 
{l_B^{-2 h_{\rm nloc}}})(\xi_{\rm cross}^{-2|\omega|}l^{h_{\rm nloc}-d} l_B^{h_{\rm nloc}-d}) \\
& =
\left(\frac{l}{l_B}\right)^{d+h_{\rm nloc}}.
\end{aligned}
\end{equation}
\end{enumerate}

\clearpage
\onecolumngrid

\section{Derivation of Eq.~\eqref{Eq:cmi_susceptibility} in the main text}
\label{app:perturbative_cmi}
  
The goal of this appendix is to provide more details on how the CMI to {leading (second) order} in $\epsilon$ for perturbed Langevin dynamics, i.e., Eq.~\eqref{Eq:cmi_susceptibility} in the main text, is obtained.
It is essentially a direct combination of the existing quadratic expansion of the CMI around a classical Markov distribution~\cite{kullback1962tests,kubkowski2021power} with the first-order McLennan--Zubarev expression for the steady state~\cite{mclennan1959statistical,zubarev1974nonequilibrium,maes2010rigorous}, both of which we briefly rederive using notation consistent with the main text.

\paragraph*{Perturbative CMI identity}
 
Let $P_0$ be Markov, i.e., $P_0[\phi]=P_0[\phi_A,\phi_B]P_0[\phi_B,\phi_C]/P_0[\phi_B]$, for the regions in Fig.~\ref{fig:geometries}(a), and perturb it as
\begin{equation}
\label{Eq:perturbed_state}
P_\epsilon[\phi]=P_0[\phi]\left[1+\epsilon V[\phi]+O(\epsilon^2)\right].
\end{equation}
Here $\epsilon V$ is the first-order relative correction to $P_0$. Probability conservation implies $\langle V\rangle_0=0$, where $\langle\cdot\rangle_0$ denotes the average over $P_0$.
To compute the corresponding CMI at leading order in $\epsilon$, it's useful to note that, for any classical distribution $P$, the CMI can be equivalently written as the relative entropy
\begin{equation}
\label{Eq:cmi_relative_entropy}
\begin{aligned}
I(A:C|B)&=D(P\|Q)=\int\mathcal D\phi\,P[\phi]\log\frac{P[\phi]}{Q[\phi]},\quad
Q[\phi]=\frac{P[\phi_A,\phi_B]\,P[\phi_B,\phi_C]}{P[\phi_B]}.
\end{aligned}
\end{equation}
Eq.~\eqref{Eq:cmi_relative_entropy} makes it apparent that the CMI measures how far $P$ is from the conditionally independent distribution built from its own marginals.
These marginals can be computed perturbatively as
\begin{equation}
\label{Eq:marginal_expansion}
\begin{aligned}
P_\epsilon[\phi_A,\phi_B]&=\int\mathcal D\phi_C\,P_\epsilon[\phi]=P_0[\phi_A,\phi_B]\left[1+\epsilon\langle V\rangle_{0|AB}\right]+O(\epsilon^2),
\end{aligned}
\end{equation}
where for any region $X$ with complement $\bar X$, we let $V_X[\phi_X]\equiv\langle V\rangle_{0|X}[\phi_X]=\int\mathcal D\phi_{\bar X}\,P_0[\phi_{\bar X}|\phi_X]\,V[\phi]$ denote the conditional average at fixed $\phi_X$. 
The analogous expressions hold for the other two marginals.
Writing $V_{AB}\equiv\langle V\rangle_{0|AB}$, and likewise $V_{BC}$ and $V_B$ for the other two marginals, we have $P_\epsilon[\phi_B,\phi_C]=P_0[\phi_B,\phi_C](1+\epsilon V_{BC})+O(\epsilon^2)$ and $P_\epsilon[\phi_B]=P_0[\phi_B](1+\epsilon V_{B})+O(\epsilon^2)$.

Inserting these into Eq.~\eqref{Eq:cmi_relative_entropy}, one finds $Q_\epsilon = P_0[1+\epsilon(V_{AB}+V_{BC}-V_B)]+O(\epsilon^2)$.  Subtracting this from $P_\epsilon$ gives
\begin{equation}
\label{Eq:residual}
P_\epsilon=Q_\epsilon + \epsilon\, P_0\, V_\times+O(\epsilon^2),\qquad
V_\times=V-V_{AB}-V_{BC}+V_{B}.
\end{equation}
It remains to expand the relative entropy in Eq.~\eqref{Eq:cmi_relative_entropy}. Letting $\delta_\epsilon\equiv P_\epsilon-Q_\epsilon$, one has
\begin{equation}
\log\frac{P_\epsilon}{Q_\epsilon}
=\log\left(1+\frac{\delta_\epsilon}{Q_\epsilon}\right)
=\frac{\delta_\epsilon}{Q_\epsilon}-\frac{\delta_\epsilon^2}{2Q_\epsilon^2}+O(\epsilon^3).
\end{equation}  
Multiplying by $P_\epsilon=Q_\epsilon+\delta_\epsilon$, retaining
terms through second order, and using
$\int\mathcal D\phi\,\delta_\epsilon=0$ (since both distributions
are normalized), we find  
\begin{equation}
\label{Eq:perturbative_cmi}
\begin{aligned}
I_\epsilon(A:C|B)=D(P_\epsilon\|Q_\epsilon)
&=\frac{1}{2}\int\mathcal D\phi\,\frac{\delta_\epsilon^2}{Q_\epsilon}+O(\epsilon^3)=\frac{\epsilon^2}{2}\,\langle V_\times^2\rangle_0+O(\epsilon^3).
\end{aligned}
\end{equation}

\paragraph*{First-order correction to the Langevin steady state}

We now derive the first-order McLennan--Zubarev expression for the steady state~\cite{mclennan1959statistical,zubarev1974nonequilibrium,maes2010rigorous}.
We consider Langevin dynamics with an added term in the deterministic part of the dynamics that need not be the gradient of any potential, so that detailed balance is broken. 
Let $\phi_i$ denote the field variables and take
\begin{equation}
\label{Eq:langevin_appendix}
\dot\phi_i=-\frac{\partial H}{\partial\phi_i}+\epsilon\, v_i(\phi)+\zeta_i,
\end{equation}
where $H$ is the Hamiltonian, $\zeta_i$ is Gaussian white noise with $\langle\zeta_i(t)\zeta_j(t')\rangle=2\delta_{ij}\delta(t-t')$, and $\epsilon v_i$ is an added term in the deterministic part of the dynamics that we keep general.
At $\epsilon=0$ the dynamics relaxes to the Gibbs distribution $P_0=e^{-H}/Z$, which for local $H$ is the Markov distribution for regions in Fig.\ref{fig:geometries}. 

To derive the perturbed probability $P_\epsilon$, we recall the Fokker--Planck equation
\begin{equation}
\label{Eq:stationary_fokker_planck}
 \sum_{i}\frac{\partial}{\partial \phi_i} \Big[ \Big( -\frac{\partial H}{\partial \phi_i} + \epsilon v_i - \frac{\partial}{\partial \phi_i}\Big)P\Big] = -\frac{\partial P}{\partial t}.
\end{equation}
For a stationary distribution, the right-hand side of Eq.~\eqref{Eq:stationary_fokker_planck} vanishes.
Write $P_\epsilon = P_0(1+\epsilon V) +O(\epsilon^2)$ with $P_0$ the Gibbs state that is the $\epsilon = 0$ stationary distribution.
Keeping the  terms of first order in $\epsilon$ gives
\begin{equation}
\begin{aligned}
0&=\sum_i\frac{\partial}{\partial\phi_i}\left[
-\frac{\partial H}{\partial\phi_i}P_0V+v_iP_0
-V\frac{\partial P_0}{\partial\phi_i} - P_0\frac{\partial V}{\partial\phi_i}\right]=\sum_i\frac{\partial}{\partial\phi_i}\left[
P_0\left(v_i-\frac{\partial V}{\partial\phi_i}\right)\right].
\end{aligned}
\end{equation}
Here we used $\partial P_0/\partial \phi_i = -(\partial H/\partial \phi_i) P_0$.  Expanding the remaining derivative once more gives
\begin{equation}
\label{Eq:steady_first_order}
\begin{aligned}
0&=P_0\sum_i\left[-v_i\frac{\partial H}{\partial\phi_i}
+\frac{\partial H}{\partial\phi_i}\frac{\partial V}{\partial\phi_i}
+\frac{\partial v_i}{\partial\phi_i}-\frac{\partial^2V}{\partial\phi_i^2}\right]=-P_0\left(\mathcal{L}^{(0)}V+\Sigma\right),
\end{aligned}
\end{equation}
where
\begin{equation}
\label{Eq:generator_and_source}
\mathcal{L}^{(0)}\equiv\sum_i\left[\frac{\partial^2}{\partial\phi_i^2}
-\frac{\partial H}{\partial\phi_i}\frac{\partial}{\partial\phi_i}\right],\quad
\Sigma\equiv\sum_i\left(v_i\frac{\partial H}{\partial\phi_i}-\frac{\partial v_i}{\partial\phi_i}\right).
\end{equation} 
In the continuum, one finds
\begin{equation}
\label{Eq:generator_continuum}
\mathcal{L}^{(0)}=\int d^dx\left[\frac{\delta^2}{\delta\phi(\mathbf{x})^2}
-\frac{\delta H}{\delta\phi(\mathbf{x})}\frac{\delta}{\delta\phi(\mathbf{x})}\right],\quad 
\Sigma[\phi]=\int d^dx\left[v(\mathbf{x})\frac{\delta H}{\delta\phi(\mathbf{x})}
-\frac{\delta v(\mathbf{x})}{\delta\phi(\mathbf{x})}\right],
\end{equation} 
with $\mathcal{L}^{(0)} V = - \Sigma$.
This implies  
\begin{equation}
\label{Eq:V_from_source}
V=(-\mathcal{L}^{(0)})^{-1}\Sigma=\int_0^\infty\!dt\;e^{t\mathcal{L}^{(0)}}\,\Sigma.
\end{equation}
Inserting this into Eq.~\eqref{Eq:perturbative_cmi} gives the CMI to order $\epsilon^2$, as written in Eq.~\eqref{Eq:cmi_susceptibility}. {Here $(-\mathcal{L}^{(0)})^{-1}$ should be understood as the pseudo-inverse on the subspace of zero-mean observables (in particular, $\langle V\rangle_0=0$).}

\section{Numerical details of the CMI scaling of the two-component model}
\label{app:two_component_exact}

We here present numerical details of the CMI scaling of the two-component model, i.e., Fig.\ref{fig:two_component_numerics} in the main text.
We will first derive the exact steady-state kernel, and then discuss how we UV regularize the kernel to calculate the CMI on a finite-size lattice.
For the reader's convenience, we repeat the Langevin equation from Eq.~\eqref{Eq:two_component}:
\begin{equation}
\begin{aligned}
\partial_t\Phi&=-(S+A)\Phi+\zeta,\qquad \Phi=(\phi,\psi)^T,\\
S&=-\begin{pmatrix}\partial_x^2+c\partial_y^2&0\\0&c\partial_x^2+\partial_y^2\end{pmatrix},
\qquad A=-\epsilon(\partial_x^3+\partial_y^3)X.
\end{aligned}
\end{equation}
Here $c>0$, $\langle\zeta_i(x,t)\zeta_j(x',t')\rangle=2\delta_{ij}\delta^2(x-x')\delta(t-t')$, and $X,Y,Z$ denote Pauli matrices acting on the two field components.
Since the dynamics is linear, the steady state is Gaussian, and its covariance $G=\langle\Phi\Phi^T\rangle=K^{-1}$ obeys the stationary Lyapunov equation
\begin{equation}
MG+GM^T=2I_2,\qquad M=S+A.
\end{equation} 
With the Fourier convention $\partial_j\mapsto ik_j$, the matrix $M=S+A$ for each nonzero momentum $\mathbf k=(k_x,k_y)$ is 
\begin{equation} 
\begin{aligned}
M(\mathbf k)&=\begin{pmatrix}a&iu\\iu&b\end{pmatrix},\quad 
a=k_x^2+ck_y^2,\ b=ck_x^2+k_y^2,\ u=\epsilon(k_x^3+k_y^3).
\end{aligned}
\end{equation}
Defining the momentum-space covariance through $\langle\Phi(\mathbf k)\Phi(\mathbf k')^T\rangle
=(2\pi)^2\delta^2(\mathbf k+\mathbf k')G(\mathbf k)$, we can write the corresponding  stationary Lyapunov equation as
\begin{equation}
\label{Eq:two_component_lyapunov_k}
M(\mathbf k)G(\mathbf k)+G(\mathbf k)M(\mathbf k)^\dagger=2I_2.
\end{equation}
Here $M(\mathbf k)^\dagger=M(-\mathbf k)^T$, so the odd-derivative term has opposite signs in $M(\mathbf k)$ and $M(\mathbf k)^\dagger$. 
Solving this $2\times2$ matrix equation gives
\begin{equation}
G(\mathbf k)=\frac{1}{(a+b)(ab+u^2)}
\begin{pmatrix}
b(a+b)+2u^2&iu(b-a)\\
-iu(b-a)&a(a+b)+2u^2
\end{pmatrix}.
\end{equation}
Inverting this covariance matrix, we obtain $K(\mathbf k)=G(\mathbf k)^{-1}$:
\begin{equation}
\label{Eq:two_component_K}
\begin{aligned}
K(\mathbf k)& =\begin{pmatrix}k_x^2+ck_y^2&0\\0&ck_x^2+k_y^2\end{pmatrix}
+
 \epsilon\frac{(c-1)\,(k_x^2-k_y^2)\,(k_x^3+k_y^3)\,
\big[\,2\epsilon(k_x^3+k_y^3)\,Z+(c+1)\,\mathbf k^2\,Y\,\big]}
{(c+1)^2\mathbf k^4+4\epsilon^2(k_x^3+k_y^3)^2}.
\end{aligned}
\end{equation}

We now discuss how we compute CMI from the exact
kernel in Eq.~\eqref{Eq:two_component_K}. 
We use a periodic $L\times L$ lattice with $L=128$ and $c=2$, in units of
the lattice spacing.  
Region $A$ is the site $(0,0)$ containing both fields,
so $l=1$;
$A\cup B$ is the square $\max(|x|,|y|)\leq l_B$ and $C$ is its complement. 
For the UV regularization, we make the replacements $k_j^2\mapsto 4 \sin^2(k_j/2),
k_x^3+k_y^3 \mapsto \sum_{j=x,y}(2\sin k_j-\sin 2k_j).$
Since the zero modes of the massless model do not have a normalizable
stationary distribution on a finite torus, we define the CMI
as the limit of an infinitesimal positive mass.
To avoid subtracting nearly equal entropies, we make use of the fact that region $A$ is small (it's a single site here) and write
\begin{equation}
\label{Eq:shared_gaussian_cmi}
I(A:C|B)=-\frac12\log\det\!\left[
I_{n_A}-K_A^{-1}K_{AC}K_C^{-1}K_{CA}\right],
\end{equation}
where $n_A = 2$ is the number of field variables in $A$ (in the long-range Ising model considered in App.\ref{app:lr_ising}, $n_A =1$). 
We construct $K_A$
and $K_{CA}$ from the inverse Fourier transform of the kernel and compute
$U\equiv K_C^{-1}K_{CA}$ by solving $K_CU=K_{CA}$
iteratively.
A similar method is used when computing CMI for the long-range Ising model in App.~\ref{app:lr_ising}.

\section{Long-range Ising model}
\label{app:lr_ising}

The goal of this appendix is to provide more details about the CMI results for the long-range Ising model discussed in the main text.

\subsection{High-temperature expansion}
\label{app:lr_ising_high_temperature}
We here use a high-temperature expansion to verify that the CMI exponent in the geometry of Fig.~\ref{fig:geometries}(b) is $\eta = 2(1+\sigma)>2$. 
Recall that the Gibbs distribution of the long-range Ising model in $d=1$ is $P_\beta(s)\propto \exp(\beta\sum_{i<j}J_{ij}s_i s_j)$, with $J_{ij}=|i-j|^{-(1+\sigma)}$.
Since $\beta\ll1$, one can directly use the perturbative CMI formula in Eq.~\eqref{Eq:perturbative_cmi} with $\epsilon=\beta$, $P_0\propto I$, and (recall $P_\epsilon = P_0(1+\epsilon V)$)
\begin{equation}
V=\sum_{\substack{i<j\\i,j\in A\cup B\cup C}}J_{ij}s_i s_j.
\end{equation}
We now compute $V_X(s_X)\equiv\langle V\rangle_{0|X}(s_X)
=\sum_{s_{\bar X}}P_0(s_{\bar X}\mid s_X)\,V(s_X,s_{\bar X})$, 
which is needed to evaluate $V_\times=V-V_{AB}-V_{BC}+V_{B}$.
Since $P_0$ is a uniform product distribution, one has $\langle s_i s_j\rangle_{0|X}=s_i s_j$ for $i,j\in X$, and the conditional average vanishes otherwise.
It follows that $V_X=\sum_{i<j,\;i,j\in X}J_{ij}s_i s_j$ for $X=A\cup B,B\cup C,B$, and thus
\begin{equation}
V_\times=V-V_{AB}-V_{BC}+V_B
=\sum_{i\in A,\;j\in C}J_{ij}s_i s_j.
\end{equation}
We next compute $\langle V_\times^2\rangle_0$ by expanding the square. Since $A$ and $C$ are disjoint, independence under $P_0$ allows us to factor the four-spin average between these two regions:
\begin{equation}
\begin{aligned}
\langle V_\times^2\rangle_0
&=\sum_{i,k\in A}\sum_{j,l\in C}J_{ij}J_{kl}\langle s_i s_j s_k s_l\rangle_0
=\sum_{i,k\in A}\sum_{j,l\in C}J_{ij}J_{kl}\langle s_i s_k\rangle_0\langle s_j s_l\rangle_0
=\sum_{i\in A,\;j\in C}J_{ij}^{\,2},
\end{aligned}
\end{equation}
where we've used $\langle s_i s_k\rangle_0=\delta_{ik}$ in the last step.
Eq.~\eqref{Eq:perturbative_cmi} therefore yields
\begin{equation}
\label{Eq:lr_ising_high_temperature_cmi}
I(A:C|B)=\frac{\beta^2}{2}\langle V_\times^2\rangle_0+O(\beta^3)
=\frac{\beta^2}{2}\sum_{i\in A,\;j\in C}\frac{1}{|i-j|^{2(1+\sigma)}}+O(\beta^3).
\end{equation}
Now, let $A$ and $C$ each contain $n$ sites, separated by $n_B\gg n$ sites. Since there are $n^2$ cross pairs, each with separation $n_B+O(n)$, we have
\begin{equation}
\label{Eq:lr_ising_high_temperature_tail}
I(A:C|B)=\frac{\beta^2 n^2}{2n_B^{2(1+\sigma)}}
\left[1+O\!\left(\frac{n}{n_B}\right)\right]+O(\beta^3).
\end{equation}

\subsection{Critical CMI in one dimension}
\label{app:lr_ising_critical_1d}
 
We now compute the CMI directly at $\mu=0$ in the geometry of Fig.~\ref{fig:geometries}(b). We restrict ourselves to the Gaussian regime $0<\sigma<1/2$, with the UV regularization $K(q)=|2\sin(q/2)|^\sigma$ of $|q|^\sigma$, where $q$ is measured in lattice units.
Let $G_m$ be the covariance matrix of $m$ consecutive sites, obtained by restricting the inverse of the full kernel to this interval, i.e., $(G_m)_{ij}=\int_{-\pi}^{\pi}\frac{dq}{2\pi}\,e^{iq(i-j)}/K(q)$ for $1\le i,j\le m$. The entropy of this Gaussian marginal is known to be $S(m)=\frac12\log[(2\pi e)^m\det G_m]$, and $\det G_m$ is known exactly~\cite{deift2014asymptotics}:
\begin{equation}
\label{Eq:lr_ising_critical_determinant}
\det G_m=\prod_{r=1}^{m}
\frac{\Gamma(r)\Gamma(r-\sigma)}{\Gamma(r-\sigma/2)^2}.
\end{equation}
Due to translational invariance, one has 
\begin{equation}
\begin{aligned}
I(A:C|B)&=2S(n_B+n)-S(n_B)-S(n_B+2n)=\frac12\log\frac{(\det G_{n_B+n})^2}{\det G_{n_B}\det G_{n_B+2n}}.
\end{aligned}
\end{equation}
Using $\Gamma(z+1)=z\Gamma(z)$, one has ${\det G_{m+1}\det G_{m-1}}/{(\det G_m)^2}=1-{\sigma^2}/{(2m-\sigma)^2}$ for $m\ge1$, with $\det G_0=1$. 
Taking the logarithm of this identity and summing over $m=n_B+i+j-1$ with $i,j=1,\ldots,n$, one finds (since the intermediate determinant terms cancel) 
\begin{equation}
\label{Eq:lr_ising_critical_cmi}
I(A:C|B)=-\frac12\sum_{i,j=1}^{n}
\log\!\left[1-\frac{\sigma^2}{\bigl(2(n_B+i+j-1)-\sigma\bigr)^2}\right].
\end{equation}
For $n_B\gg n$, expanding each logarithm immediately gives
\begin{equation}
I(A:C|B)=\frac{\sigma^2}{8}\left(\frac{n}{n_B}\right)^2
\left[1+O\!\left(\frac{n}{n_B}\right)\right],
\end{equation}
and thus $ \eta=2$.

\subsection{Numerical verification in the Gaussian regime}
 
We now provide numerical support for Eqs.~\eqref{Eq:large_y_regime_b} and \eqref{Eq:large_y_regime_c}
in $d=2$ for $\sigma=0.6$ and $0.8$, with the lattice quadratic kernel 
\begin{equation}
\label{Eq:lr_gaussian_lattice}
K(\mathbf q)=\mu+\left[4\sum_{j=x,y}\sin^2(q_j/2)\right]^{\sigma/2}.
\end{equation}
We take $A$ to be the origin ($l=1$) and $C=\{(x,y):l_B<\max(|x|,|y|)\le8l_B\}$. We use a periodic Fourier grid of side $L=32l_B$ and fix the field to zero wherever $\max(|x|,|y|)>8l_B$. Doubling both $L$ and the outer cutoff changes the sampled CMI values only slightly.  
We use the same method discussed in  App.\ref{app:two_component_exact} to compute the corresponding CMI.

\begin{figure}[t]
\centering
\includegraphics[width=0.48\textwidth]{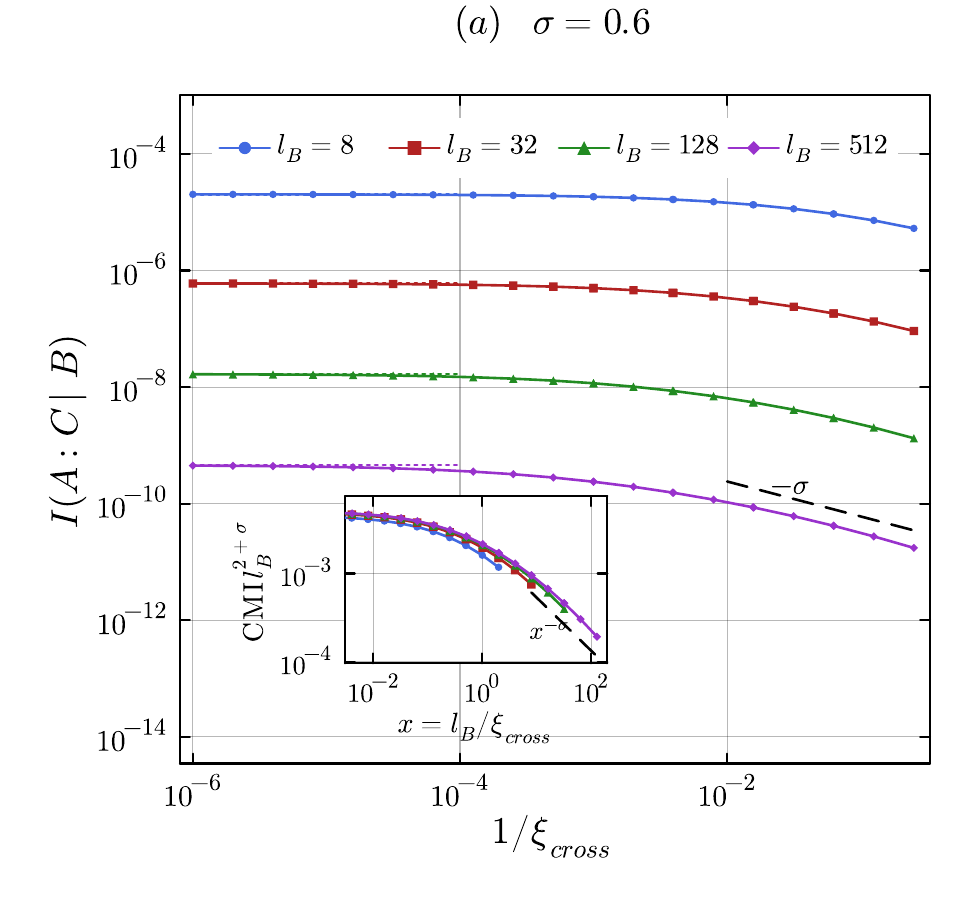}\hfill
\includegraphics[width=0.48\textwidth]{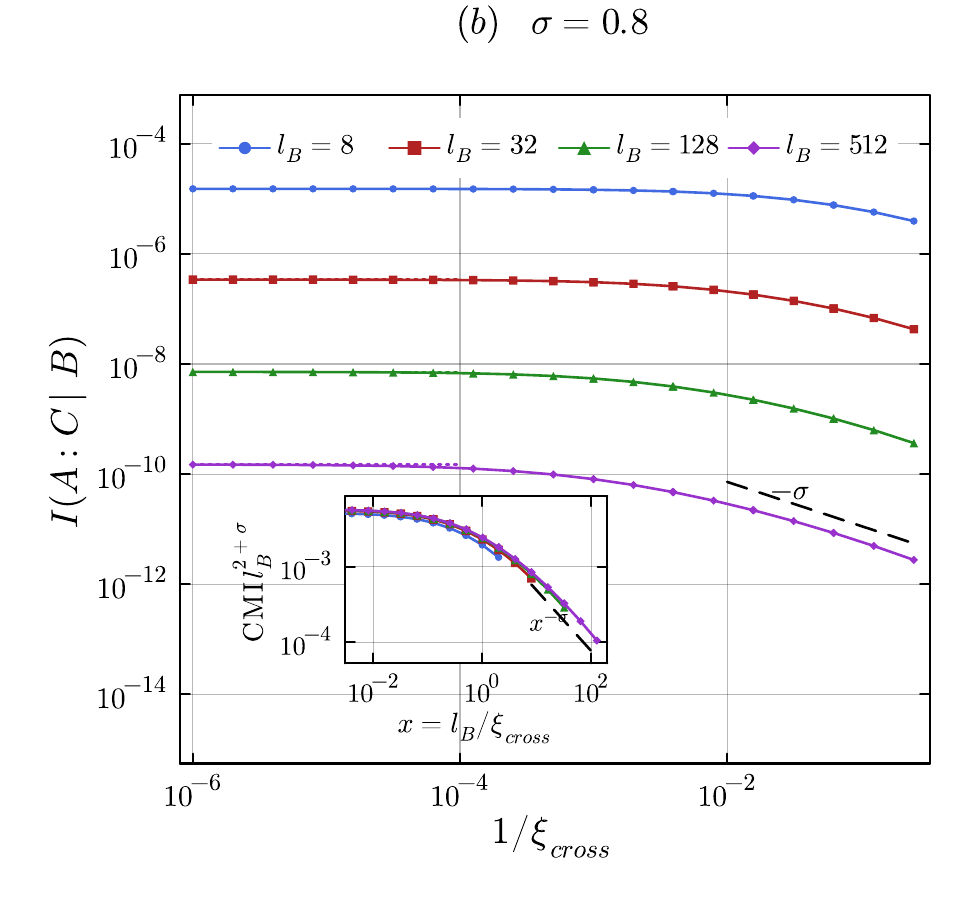}
\caption{CMI of the long-range Gaussian model in $d=2$ with
(a) $\sigma=0.6$ and (b) $\sigma=0.8$, for a single-site $A$.
Main panels: $I(A:C|B)$ versus $1/\xi_{\rm cross}$ at fixed $l_B$;
dotted lines mark the $\mu=0$ limits.
Insets: CMI$\,l_B^{2+\sigma}$ versus $l_B/\xi_{\rm cross}$.}
\label{fig:lr_numerics}
\end{figure}

Recall that Eqs.~\eqref{Eq:large_y_regime_b} and \eqref{Eq:large_y_regime_c} predict (with $h_{\rm loc} = 0$, $h_{\rm nloc} = \sigma$)
\begin{equation}
\label{Eq:app_scaling}
I(A:C|B)\,l_B^{2+\sigma}\sim
\begin{cases}
\text{const.},&1\ll l_B\ll\xi_{\rm cross},\\
(l_B/\xi_{\rm cross})^{-\sigma},&1\ll\xi_{\rm cross}\ll l_B.
\end{cases}
\end{equation}
Fig.~\ref{fig:lr_numerics} shows the crossover between these regimes.
The main panels plot CMI against $1/\xi_{\rm cross}$ (where $\xi_{\rm cross}=\mu^{-1/\sigma}$) at fixed $l_B$. 
The insets show the data collapse that supports Eq.\eqref{Eq:app_scaling}.

\end{document}